\documentclass[%
 reprint,
superscriptaddress,
 amsmath,amssymb,
 aps,
]{revtex4-2}
\usepackage{color}
\usepackage{graphicx}
\usepackage{dcolumn}
\usepackage{bm}
\usepackage[export]{adjustbox} 

\begin{document}


\title{How do polymers stretch in capillary-driven extensional flows? }

\author{Vincenzo Calabrese}
\email{vincenzo.calabrese@oist.jp}
\affiliation{Micro/Bio/Nanofluidics Unit, Okinawa Institute of Science and Technology Graduate University\\1919-1 Tancha, Onna-son, Okinawa 904-0495, Japan}%

\author{Amy Q. Shen}
\affiliation{Micro/Bio/Nanofluidics Unit, Okinawa Institute of Science and Technology Graduate University\\1919-1 Tancha, Onna-son, Okinawa 904-0495, Japan}%
\author{Simon J. Haward}
\email{simon.haward@oist.jp}
\affiliation{Micro/Bio/Nanofluidics Unit, Okinawa Institute of Science and Technology Graduate University\\1919-1 Tancha, Onna-son, Okinawa 904-0495, Japan}%

\date{\today}

\begin{abstract}
Measurements of the capillary-driven thinning and breakup of fluid filaments are widely used to extract extensional rheological properties of complex materials. For viscoelastic (e.g., polymeric) fluids, the determination of the longest relaxation time depends on several assumptions concerning the polymeric response to the flow that are derived from constitutive models. Our capillary thinning experiments using polymeric fluids with a wide range of extensibility, suggest that these assumptions are likely only valid for highly extensible polymers but do not hold in general. For polymers with relatively low extensibility, such as polyelectrolytes in salt-free media, conventional extrapolation of the longest relaxation time from capillary thinning techniques leads to a significant underestimation.
\end{abstract}

\maketitle

\begin{figure}
   \centering
    \includegraphics[width=0.46\textwidth]{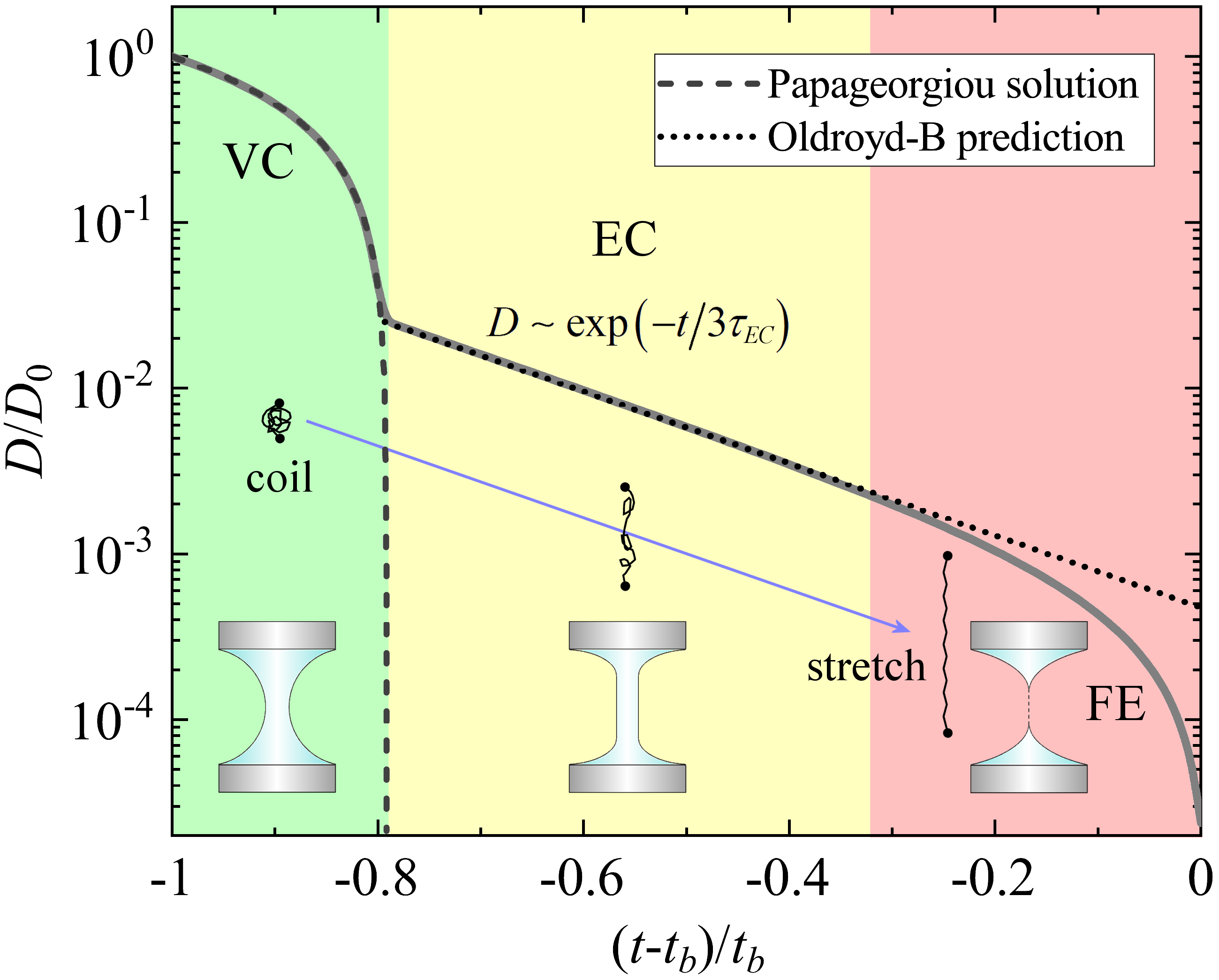}
 \caption{A schematic representation of a typical capillary-thinning (SRM) experiment performed with a viscoelastic fluid sample. See text for description.}
  \label{Fig:sch}
\end{figure}

Measuring the diameter $D$ as a function of time $t$ at the midpoint of a fluid filament as it thins under the action of capillarity has become a common method for estimating the extensional rheological properties of polymeric fluids. There are many methods for initiating the capillary-driven thinning of a liquid bridge, the subsequent dynamics of which can all be analyzed in a roughly similar way to obtain rheometric data.\cite{Anna2001,Campo2010,Nelson2011,Bhattacharjee2011,Dinic2015,Keshavarz2015,Sharma2015,Mathues2018,Rajesh2022,Gaillard2023,vadillo2012microsecond} We focus on a technique termed the slow retraction method (SRM),\cite{Campo2010} in which a fluid sample is loaded between coaxial circular plates initially separated vertically by a small distance ($\sim 1~\text{mm}$). The bottom plate is gradually displaced vertically downwards until the liquid bridge between the plates (now with diameter $D=D_0$) becomes unstable to capillary forces and begins to break up. For a Newtonian fluid in the absence of inertia, the liquid bridge attains an hourglass shape and undergoes viscocapillary (VC) thinning during which the filament diameter decays linearly with time at a rate that depends on the surface tension $\sigma$ and the shear viscosity $\eta$ (as shown by Papageorgiou, see Fig.~\ref{Fig:sch}).\cite{Papageorgiou1995,McKinley2000} For a polymeric fluid, the (inertialess) thinning of the liquid bridge at short times is Newtonian-like (Fig.~\ref{Fig:sch}) and commonly associated with the initially coiled state of the polymer. Importantly, in the VC regime the extension rate in the fluid neck (estimated by $\dot\varepsilon(t) = -[2/D(t)] \text{d}D(t)/\text{d}(t)$) increases with time. 

According to the prediction of the Oldroyd-B constitutive model, which considers a polymer as being infinitely extensible, as $\dot\varepsilon$ increases beyond a critical value $\dot\varepsilon_c$, polymers in the fluid filament begin to deform from their equilibrium coils. As the polymer stretches, the elastic stress grows and begins to dominate over the viscous stress. Consequently,  the thinning behavior of the filament diverges from the Newtonian case, leading to the onset of an elastocapillary (EC) thinning regime (Fig.~\ref{Fig:sch}).\cite{Bazilevsky1990,Entov1997,Anna2001,clasen2006dilute}
Through the EC regime, the filament develops into a slender columnar shape, the diameter of which decays exponentially in time with the consequence that $\dot\varepsilon$ becomes constant with a value $\dot\varepsilon_{EC}$. However, there is no control over the value of $\dot\varepsilon_{EC}$, which is self-selected by the fluid via the force balance. Fitting the exponential decay in the EC regime with the form
\begin{equation}
D \sim \exp({-t/3\tau_{EC}}),
\label{eqn:expEC}
\end{equation}
is a standard procedure used to extract a characteristic time constant for the fluid (here denoted as $\tau_{EC}$). This time constant is often considered to be a fundamental material property, commonly described as the longest relaxation time ($\tau$) of the polymer.\cite{Gaillard2023} It can be shown that $\dot\varepsilon_{EC} \tau_{EC} = 2/3$,\cite{gier2012visualization} thus it is assumed that the Weissenberg number in the EC regime $Wi_{EC} = \dot\varepsilon_{\text{EC}} \tau = 2/3$ (i.e., $Wi_{EC}>\dot\varepsilon_c \tau =0.5$) and hence that polymers will become stretched.\cite{DeGennes1974,Hinch1977,Larson1989,Perkins1997} 

For the Oldroyd-B model, which considers a polymer as being infinitely extensible, the EC regime is predicted to persist indefinitely.  However for real polymeric fluids the filament thinning eventually becomes super exponential and the filament undergoes pinch off at time $t=t_b$ (Fig.~\ref{Fig:sch}). The finite $t_b$ can be predicted by including finite extensibility in the constitutive model, hence the labelling of this terminal ``FE'' regime.\cite{Entov1997,Anna2001,Wagner2015} The finitely-extensible non-linear elastic (FENE-P) dumbbell model predicts that, as polymers approach full extension, they begin to behave as rigid rods. In the FE regime, the fluid exhibits Newtonian behavior, with the filament thinning linearly in time, similar to the VC regime but with a higher (extensional) viscosity. \cite{McKinley2005,Miller2009,Dinic2019}

The Oldroyd-B and FENE-P models predict an intimate connection between the macromolecular and the filament thinning dynamics, which was largely borne out by early capillary thinning measurements performed on solutions of highly flexible polymers with well defined molecular parameters.\cite{Liang1994,Anna2001} However, the direct assessment of polymer dynamics during capillary thinning remains untested due to considerable experimental difficulties. This may explain why predictions from the Oldroyd-B and FENE models have been broadly applied to interpret capillary thinning data obtained from almost any type of polymer in any solvent.\cite{Plog2005,Duxenneuner2008,Haward2012b,Jimenez2018,Walter2019,Rosello2019,Jimenez2020,Martinez2021,Lauser2021,Jimenez2022,Zhang2022,Soetrisno2023,morozova2018extensional}

%

Of particular relevance, we refer to solutions of fluorescently-labelled DNA (considered as a model polymer for molecular rheology)\cite{latinwo2011model}, the dynamics of which have been directly observed under high fluid strains in steady extensional flows in cross-slot devices at precisely controlled values of $\dot\varepsilon$.\cite{Schroeder2003,Schroeder2018} Such experiments provide a clear and unambiguous value for the longest relaxation time $\tau=0.5/\dot\varepsilon_c  \sim \mathcal{O}(\text{s})$, where $\dot\varepsilon_c$ is the extension rate beyond which the DNA molecules are seen to accumulate strain.\cite{Perkins1997,Smith1998,Schroeder2003,Schroeder2018} However, capillary thinning experiments on similar fluids show weakly pronounced EC regimes suggestive of characteristic times $\tau_{EC} \sim \mathcal{O}(\text{ms})$,\cite{Hsiao2017} i.e., $\tau_{EC} \ll \tau$, an apparent inconsistency that strongly motivated the current work.



\begin{figure}[t]
   \centering
    \includegraphics[width=0.51\textwidth]{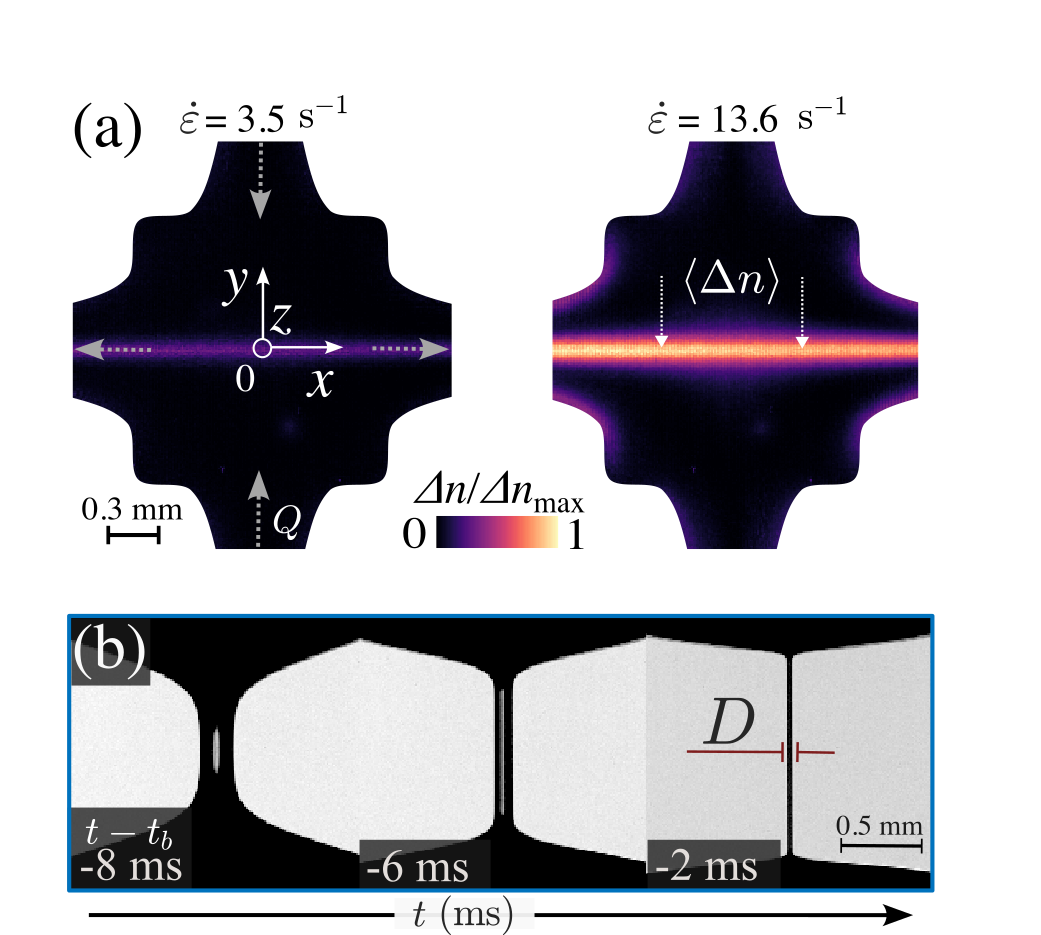}
 \caption{(a) Birefringence $\Delta n$ (normalized by its maximum value, $\Delta n_{\text{max}}$) in a $\lambda$-DNA solution at $c=1.1c^*$,  flowing in a microfluidic OSCER geometry at two well-defined extension rates. The characteristic dimensions of the channel are the width of the channel arm $W$ and the height $H$ along the $z$-axis corresponding to the optical path. The flow in each arm is driven by a syringe pump at a controlled volumetric rate $Q$. The flow scheme and coordinate system are indicated in the left-hand image. The OSCER generates a planar extensional flow that is shear-free around the stagnation point $x=y=0$~ mm, the extension being along the $x$ direction. Birefringence is measured with the OSCER device on an upright microscope and by using a Photron Crysta PI-1P high speed polarization imaging camera. (b) Snapshots at different moments during an SRM experiment with the same $\lambda$-DNA solution used in part (a). SRM is implemented on a commercial capillary breakup extensional rheometer (CaBER 1, Thermo Haake) fitted with end plates of diameter $D_p=6$~mm, and imaging is performed using a Phantom Miro 310 high speed camera.  
}
  \label{Fig:snap}
\end{figure}

In this Letter, we provide insight into the polymer dynamics occurring during the capillary thinning of polymeric fluids, highlighting the strengths and limitations of filament thinning techniques for extensional rheometry. We provide a generalized perspective by analyzing a wide variety of polymers with different extensibility $L = l_c/\sqrt{\langle R^2 \rangle}$, where $l_c$ is the contour length and $\langle R^2 \rangle$ is the mean-squared end-to-end length of the polymer at rest. For each polymeric solution, we perform birefringence measurements in an optimized form of the microfluidic cross-slot geometry (e.g., see Fig.~\ref{Fig:snap}(a))\cite{Haward2012,Haward2016b} to determine a value of $\tau$ associated with the longest time scale at which polymer chains exhibit anisotropy. We also examine the response in an SRM experiment (e.g., Fig.~\ref{Fig:snap}(b)) to obtain a value for $\tau_{EC}$. Our findings reveal a clear correlation between the ratio $\tau_{EC}/\tau$ and $L$, showing $\tau_{EC} \approx \tau$ for large extensibility, but a strong decrease in $\tau_{EC}/\tau$ as $L$ becomes small. By integrating the strain rates measured over time during capillary thinning experiments, we show that for high $L$, the estimates of macromolecular strain are in line with the predictions of the models. However, for polymers with lower $L$ a significant degree of extension (even full extension) is likely even before the onset of the EC regime, a dynamic that is not captured by the existing models.
Fig.~\ref{Fig:snap} displays exemplary results for a $\lambda$-DNA solution in a viscous buffer solution (Tris-EDTA, 5 mM NaCl, and sucrose with viscosity $\eta_s=4.3$~mPa~s) in an optimized-shape cross-slot extensional rheometer (OSCER, Fig.~\ref{Fig:snap}(a)) and during an SRM experiment (Fig.~\ref{Fig:snap}(b)). The $\lambda$-DNA has extensibility $L\approx 12$ (see ESI) and is prepared at a concentration $c=1.1c^*$ where $c^*=40$~\textmu g/mL is the overlap concentration.\cite{Hsiao2017}  Fig.~\ref{Fig:snap}(a) shows the birefringence $\Delta n$ at two distinct extension rates $\dot\varepsilon \approx 0.2 Q/ (W^2 H)$, where $Q$ is the volumetric flow rate and $H$, $W$ are the height and width of the channel, respectively.\cite{Haward2012} Above a critical extension rate, $\dot\varepsilon_c$, the polymer starts to unravel and a birefringent strand appears along the $x$-axis of the device. With increasing $\dot\varepsilon$ the intensity of the birefringent strand increases, indicating pronounced DNA stretching. 
In the SRM experiment (Fig.~\ref{Fig:snap}(b)), we utilize high-speed imaging to monitor the filament as it undergoes self-thinning over time $t$ until it reaches pinch-off at time $t=t_b$. For $t-t_b >-8$~ms, the filament develops a cylindrical shape, characteristic of the EC regime encountered for polymer solutions.

\begin{figure}[ht!]
\includegraphics[width=7cm]{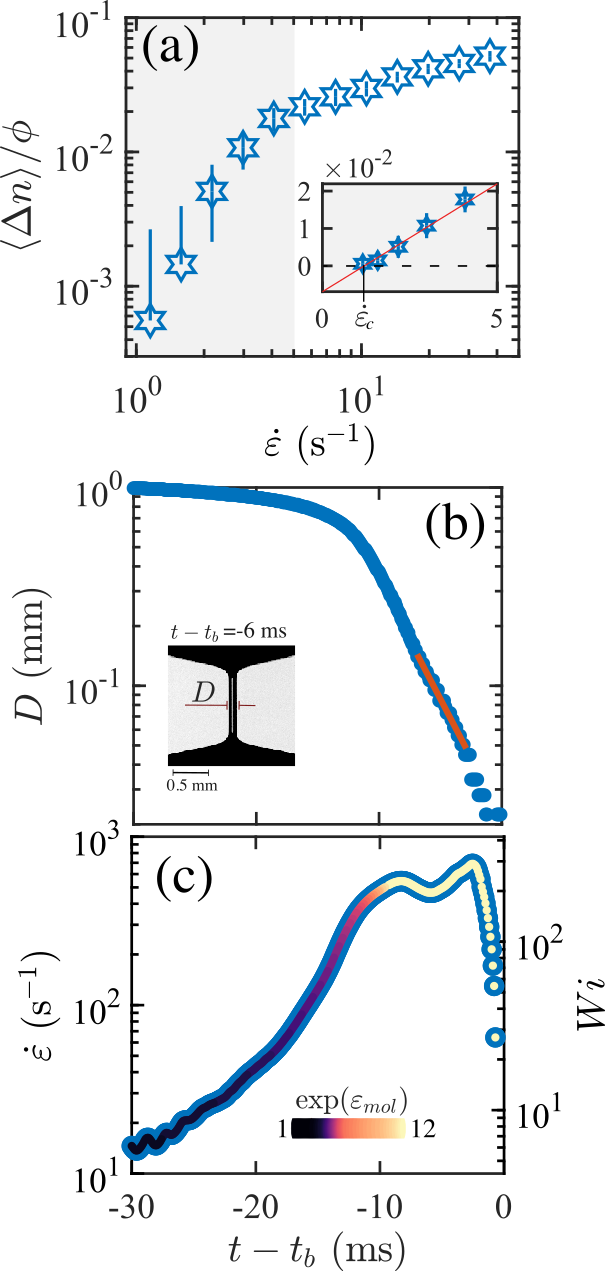}
\caption{Results from (a) microfluidic and (b, c) SRM experiment for a $\lambda$-DNA solution at $c=1.1c^*$. (a) The time-averaged intensity of the birefringent strand $\langle \Delta n \rangle$, normalized by the polymer mass fraction $\phi$, as a function of the extension rate $\dot\varepsilon$. The inset shows the lin-lin plot of $\langle \Delta n\rangle/ \phi$ \textit{vs.} $\dot\varepsilon$ with the linear fitting at $\langle \Delta n\rangle/ \phi \rightarrow 0$ used to estimate the extension rate at the onset of birefringence ($\dot\varepsilon_c$), and the relaxation time $\tau=0.5/\dot\varepsilon_c$. (b) Filament diameter $D$ as a function of time $t-t_b$. The line is the fitting to eqn.~\ref{eqn:expEC}. (c) The extension rate at the neck of the filament ($\dot\varepsilon(t) = -[2/D(t)] \text{d}D(t)/\text{d}(t)$) and the respective $Wi=\dot\varepsilon \tau$ as a function of time $t-t_b$. The color scale indicates the accumulated macromolecular strain, $\exp(\varepsilon_{mol})$. }
  \label{fgr:DNA}
\end{figure}

From the OSCER experiment, we retrieve the time-averaged intensity of the birefringent strand, $\langle \Delta n \rangle$ (i.e., at $y=0$~mm for $-1.5 W \lesssim x \lesssim 1.5W$, see Fig.\ref{Fig:snap}(a)) as a function of $\dot\varepsilon$, as shown normalized by the polymer mass fraction ($\phi$) in Fig.~\ref{fgr:DNA}(a). Linear extrapolation of the birefringence data for small $\dot\varepsilon$ (red line, inset Fig.~\ref{fgr:DNA}(a)), provides a precise value of $\dot\varepsilon_c = 1.2~\text{s}^{-1}$ for the onset of orientation of the polymer chains. Since at the onset of birefringence we expect the Weissenberg number $Wi=\dot\varepsilon \tau \approx 0.5$,\cite{Perkins1997,Smith1998,Schroeder2003,Schroeder2018} the relaxation time can be estimated as $\tau \approx 0.5/\dot\varepsilon_c$, hence $\tau=0.42$~s. Accounting for the effects of solvent viscosity and $\lambda$ -DNA concentration,\cite{zhou2018dynamically} expectations based on single molecule imaging experiments \cite{Smith1998,smith1999single,teixeira2005shear} and an estimate of the Rouse relaxation time based on an experimental intrinsic viscosity measurement \cite{shrewsbury2001effect}, suggest a value of $\tau$ in the range $0.25\lesssim\tau\lesssim 0.52$~s, consistent with the current result. 

The value of $\tau$ retrieved from the microfluidic experiment suggests an elastocapillary number in the SRM experiment $Ec=2 \tau \sigma / \eta_s D_p \gg 1 $, where $D_p = 6$~mm is the diameter of the end-plates. This suggests that the elastic stress from the polymer should be sufficient to dominate over the viscous stress of the solvent, provided that the polymer concentration is sufficiently high.\cite{clasen2006dilute}  Following  Clasen et al., the concentration of the tested $\lambda$-DNA solution is estimated to be significantly greater than the minimum concentration required to observe elastic effects on the thinning dynamics  (i.e, $c>c_{min}=1.8$~\textmu g/mL, see ESI).\cite{clasen2006dilute} The $\lambda$-DNA solution shows a narrow EC regime where $D$ decreases exponentially over time ($-8\lesssim t-t_b \lesssim-3$~ms, Fig.~\ref{fgr:DNA}(b)). The presence of the EC regime is also clear from the plateau-like region in $\dot\varepsilon$ \textit{vs} $t-t_b$ (Fig.~\ref{fgr:DNA}(c)). By fitting eqn.~\ref{eqn:expEC} to the EC region, we retrieve $\tau_{EC}=1.1$~ms, a value much smaller than $\tau$ obtained from the birefringence experiment ($\tau_{EC} \approx \tau/380$). Consequently, the steady value of $\dot\varepsilon_{EC}$ sets $Wi_{EC}=\dot\varepsilon \tau \approx 250$, in stark contrast to the value of $Wi_{EC}=2/3$ predicted by the Oldroyd-B and FENE-P models. 

In general, if coiled macromolecules are exposed to a persistent extensional flow with $Wi=\dot\varepsilon \tau \gtrsim 0.5$,  then macromolecular strain should accumulate over time. We roughly estimate, similarly to Ref.~\citenum{Perkins1997}, the accumulated macromolecular strain, in the SRM experiment as $\varepsilon_{mol}(t)=\int_{-t_b}^{t} (\dot\varepsilon-\dot\varepsilon_c) \text{d}t$ for $Wi\geq 0.5$. Since fluid elements accumulate strain exponentially in extensional flows, we expect (assuming affine deformation of the polymer for any $Wi\geq 0.5$) that the $\lambda$-DNA reaches a fully stretched state when $\exp(\varepsilon_{mol})\gtrsim L \approx 12$. Based on our estimate (shown by the color-coded data points in Fig.~\ref{fgr:DNA}(c)), the value of $\exp(\varepsilon_{mol}) =12$ is reached at $t-t_b \approx -10$~ms, suggesting that the polymer is likely to be highly stretched even before the onset of the EC regime, a scenario that is not predicted by the Oldroyd-B model.  This is consistent with a criterion proposed by Campo-Dea\~no and Clasen, based on the finite extensibility of the polymer, which suggests that the elastic stress generated by the $\lambda$-DNA at the onset of stretching should be insufficient to induce an EC regime for $c< c_{low} = 470$~\textmu g/mL (see ESI for estimates).

\begin{figure}[t!]
\includegraphics[width=7cm]{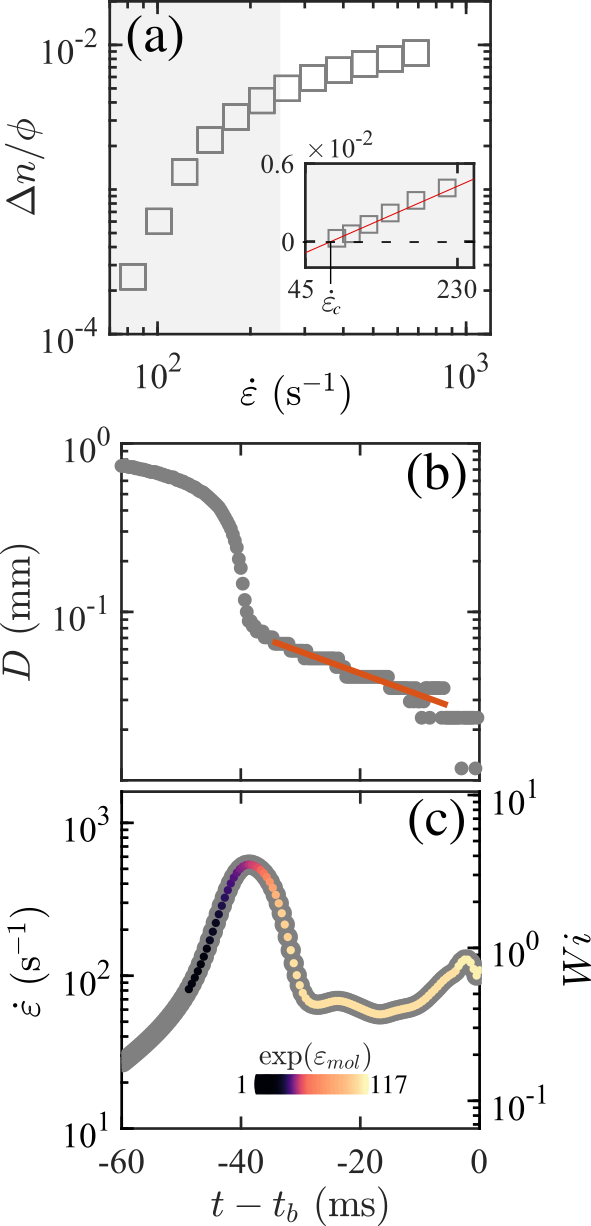}
\caption{Results from (a) microfluidic and (b, c) SRM experiments for a PS solution ($\text{Mw}=7$~MDa) at $c=0.025c^*$. (a) Time-averaged intensity of the birefringent strand $\langle \Delta n \rangle$, normalized by the mass fraction $\phi$, as a function of the extension rate $\dot\varepsilon$. The inset shows the lin-lin plot of $\langle \Delta n\rangle/ \phi$ \textit{vs.} $\dot\varepsilon$ with the linear fitting at $\langle \Delta n\rangle/ \phi \rightarrow 0$ used to estimate the extension rate at the onset of birefringence ($\dot\varepsilon_c$), and the relaxation time $\tau=0.5/\dot\varepsilon_c$. (b) Filament diameter $D$ as a function of time $t-t_b$. The line is the the fitting to eqn.~\ref{eqn:expEC}. (c) Extension rate at the neck of the filament ($\dot\varepsilon(t) = -[2/D(t)] \text{d}D(t)/\text{d}(t)$) and the respective $Wi=\dot\varepsilon \tau$ as a function of time $t-t_b$. The color scale indicates the accumulated macromolecular strain, $\exp(\varepsilon_{mol})$.}
  \label{fgr:Fig3}
\end{figure}

We compare our analysis for the $\lambda$-DNA with a significantly more extensible polymer that aligns more closely with the assumptions of the Oldroyd-B model, namely polystyrene (PS) with molecular weight $\text{Mw}=7$~MDa in a $\theta$ solvent (dioctyl phthalate, DOP) and a resulting $L\approx 117$. The dilute PS solution ($c=0.025c^*>c_{low}>c_{min}$, $Ec\approx 1$) shows an onset of birefringence at $\dot\varepsilon_c= 76~\text{s}^{-1}$ (Fig.~\ref{fgr:Fig3}(a)), from which a relaxation time $\tau=6.5$~ms is estimated. 
In the SRM experiment (Fig.~\ref{fgr:Fig3}(b)), the PS solution shows a clear EC thinning regime, yielding $\tau_{EC}= 10$~ms (eqn.~\ref{eqn:expEC}), in reasonable agreement with the value $\tau=6.5$~ms obtained from the birefringence experiment. As such, the steady state region of $\dot\varepsilon_{EC}$ yields $Wi_{EC}\approx 0.5$, close to the $Wi_{EC}=2/3$ expected from the Oldroyd-B model (Fig.~\ref{fgr:Fig3}(c)). An estimate of the macromolecular strain in this case (see the color-coded data points in Fig.~\ref{fgr:Fig3}(c)) indicates that the transition into the EC regime (i.e., beyond the peak in $\dot\varepsilon$) roughly coincides with the initial unraveling of the polymers (i.e., $\exp(\varepsilon_{mol})>1$) and that maximum extension is approached (i.e., $\exp(\varepsilon_{mol}) \rightarrow L$) towards the end of the experiment at $t=t_b$. 
 In this case, our roughly estimated macromolecular dynamics are in broad consistency with the models  and with a criterion proposed by Campo-Dea\~no and Clasen suggesting that  the stress generated by the onset of polymer stretching should be sufficient to induce an EC regime (i.e., $c> c_{low}>c_{min}$, see ESI), resulting in consistency between $\tau_{EC}$ and the longest relaxation time $\tau$.\cite{Campo2010}

While both PS and $\lambda$-DNA share the presence of the EC regime at the macroscopic fluid level, the underlying polymer dynamics occurring during the EC regime are distinct.  We use the ratio $\tau_{EC}/\tau$ to capture the polymer dynamic that occurs in capillary thinning for polymers with a wide range of extensibility, $L$. For stiffer polymers, we use hyaluronic acid (HA) and carboxymethyl cellulose (CMC) in deionized water, along with monodisperse $\lambda$ -DNA (discussed above) and polydisperse calf thymus DNA (ct-DNA) with average contour length comparable to $\lambda$-DNA. For more flexible polymers we use PS in DOP, polyethylene oxide (PEO) in aqueous polyethylene glycol (PEG, $\text{Mw} = 8$~kDa) solution, and hyaluronic acid in a 1.5~M~NaCl aqueous solution (HA\textsubscript{NaCl}), all with a range of $\text{Mw}$. Polymer concentrations, typically between the dilute and semi-dilute regimes, are chosen to be as low as possible but still sufficient to probe the EC region in the SRM\cite{clasen2006dilute} experiment and have detectable birefringence in the microfluidic OSCER device (see ESI for detailed specifications of the tested polymer samples). Note that we tested several fluids by SRM using end plates of various diameter, with no systematic variation in the obtained values of $\tau_{EC}$ (data shown in ESI).\cite{Gaillard2023}

\begin{figure*}
\includegraphics[max size={\textwidth}]{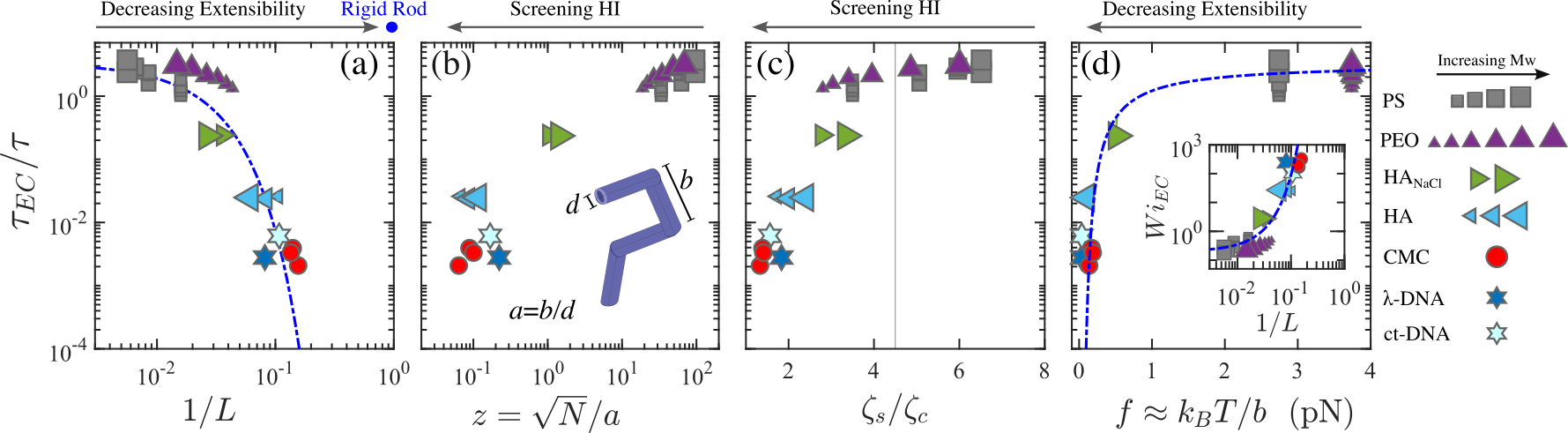}
\caption{Ratio $\tau_{EC}/\tau$ as a function of (a) $1/L$, (b) $z=\sqrt{N}/a$, (c) $\zeta_s/\zeta_c$ and (d) $f \approx k_BT/b$ for different polymer solutions. Insert in (d) shows the Weissenberg number in the EC regime $Wi_{EC}$ as a function of $1/L$. The dashed lines, given as reference functions, are (a) $\tau_{EC}/\tau=3.3~\text{exp}(55/(1-L))$, (d) $\tau_{EC}/\tau=3.3~ \text{exp}(-1/f)$ and (d, insert) $Wi_{EC}=0.2~\text{exp}(55/(L-1))$. For a given polymer, the increasing $\text{Mw}$ is depicted by a an increasing symbol size. For PS and CMC, multiple data points indicate different concentrations.  }
  \label{fgr:Fig4}
\end{figure*}

In Fig.~\ref{fgr:Fig4}(a) we plot the ratio $\tau_{EC}/\tau$ as a function of $1/L$. For the most flexible PS and PEO polymers with $L \gtrsim 30$, $\tau_{EC}/\tau~\sim \mathcal{O}(1)$. We note that this result holds despite the widely different volatility of the respective solvents.\cite{colby2023fiber}  For more rigid polymers with $L\lesssim 20$, $\tau_{EC}/\tau$ decreases significantly below unity. For the stiffest DNA and CMC, $\tau_{EC}/\tau ~\sim \mathcal{O}( 10^{-2})$. At least qualitatively, it is clear that as $1/L$ increases, $\tau_{EC}/\tau$ decreases. Possible reasons for the extent of data scatter could be linked to differences in sample polydispersity, the difficulty in estimating $L$, and the inherent molecular differences between the tested polymers.

We interpret the trend of $\tau_{EC}/\tau$ as a function of polymer extensibility as follows. In the birefringence experiment, we measure the longest relaxation time $\tau$, associated with the time scale ($0.5/\dot\varepsilon_c$) at which the orientation of the polymer chains occurs.\cite{Perkins1997} On the other hand, the capillary thinning experiment relies on the stretching polymer to induce sufficient elastic stress in order to dominate the viscous stress and initiate an EC thinning regime.\cite{clasen2006dilute} We may consider $\tau_{EC}$ as a time scale associated with the onset of \textit{measurable} elastic stresses (by which we mean sufficient to cause a deviation from Newtonian-like thinning).\cite{clasen2006dilute,Wagner2015} Evidently, $\tau$ and $\tau_{EC}$ are not, in general, the same.

For highly flexible polymers such as PS and PEO, $\tau_{EC}/\tau~\sim \mathcal{O}(1)$ (Fig.~\ref{fgr:Fig4}), indicating that the time scale of chain orientation corresponds to the time scale at which elastic stresses becomes detectable in a capillary thinning experiment. Note that for flexible polymer solutions, $\tau_{EC}$ can not be significantly greater than $\tau$ as the polymer must orient and stretch in order for the elastic stress to develop.

With decreasing polymer extensibility, $\tau_{EC}$ decreases relative to $\tau$. This suggests that chain orientation occurs at a longer time scale than that required for observing elastic effects in the capillary thinning experiment. It is instructive to consider the limiting case of rigid rod-like polymers ($L = 1$) that orient in flow at an extension rate $\dot\varepsilon_c$ sufficient to overcome rotational diffusion and thus have finite $\tau>0$.\cite{doi1988Book,Lang2019,Calabrese2021,tsentalovich2016relationship} However, since rigid rod-like polymers have negligible entropic elasticity,\cite{rubinstein2003polymer} we expect $\tau_{EC}\rightarrow 0$ and the total absence of the EC regime (see, e.g., Refs.~\citenum{lundahl2018shear,lang2019effects,calabrese2022microstructural}). Therefore, we expect two natural asymptotes in the plot of $\tau_{EC}/\tau$ versus $1/L$: i.e., $\tau_{EC}/\tau \rightarrow \sim1$ as $1/L \rightarrow 0$, and $\tau_{EC}/\tau \rightarrow 0$ as $1/L \rightarrow 1$. This is roughly captured by the empirically-determined function $\tau_{EC}/ \tau = 3.3~\text{exp}(55/(1-L))$ (dashed line in Fig.\ref{fgr:Fig4}(a)).

In Fig.~\ref{fgr:Fig4}(b) we plot the ratio $\tau_{EC}/\tau$ as a function of a chain interaction parameter $z=\sqrt{N}/a$ (where $N$ is the number, and $a=b/d$ is the aspect ratio of each Kuhn segment in the polymer chain, where $b$ and $d$ are the length and thickness, respectively). For large $\text{Mw}$ polymers with  relatively small Kuhn segments, $z > 1$ indicates strong intramolecular hydrodynamic interactions (HI).\cite{latinwo2011model} The $\lambda$-DNA and PS ($\text{Mw} = 7~\text{MDa}$) previously described have estimated values of $z=0.45$ and $z=60$, respectively. This indicates that the $\lambda$-DNA coil adopts a free-draining configuration with screened HI that keep the polymer segments hydrodynamically unshielded.\cite{latinwo2011model}  On the contrary, the PS with $z\gg 1$ forms a non-free-draining coil with hydrodynamically shielded polymer segments. 

The ratio of the hydrodynamic drag between the stretched ($\zeta_s$) and coiled ($\zeta_c$) conformations of a polymer, estimated as $\zeta_s/\zeta_c \approx (N^2 a )^{1/5} / \text{ln}(Na)$, gauges the importance of HI to the stretching dynamics.\cite{latinwo2011model,Schroeder2003,hsieh2005prediction}  Based on experiments and simulations,  $\zeta_s/\zeta_c \approx 4.5$ sets the threshold value above which HI become dominant and the polymer segments are hydrodynamically shielded. The plot of $\tau_{EC}/\tau$ as a function of $\zeta_s/\zeta_c$ shows that the plateau $\tau_{EC}/\tau~\sim \mathcal{O}(1)$  is reached for $\zeta_s/\zeta_c \gtrsim 4$ (Fig.~\ref{fgr:Fig4}(c)). Based on the analysis of the parameters $z$ and $\zeta_s/\zeta_c$, we hypothesise that stiffer polymers with screened HI (i.e., $z < 1$ and  $\zeta_s/\zeta_c \lesssim 4.5$) leads to an easier chain orientation even before the EC sets in, resulting in $\tau_{EC}<\tau$. Additionally, stiffer polymers exert less stress into the flow. This is evidenced by computing the chain tension above which the polymer behaves as a nonlinear spring, $f \approx k_BT/b$, where $k_B$ is the Boltzmann constant and $T$ is the temperature (Fig.~\ref{fgr:Fig4}(d)).\cite{rubinstein2003polymer} For the more flexible chains (PS and PEO) with significant HI this force is several pico Newtons per chain, but for the stiffer chains (e.g., CMC and DNA) $f$ is extremely low.  

We suggest that for highly extensible polymers with strong HI (i.e., $z\gg 1$ and $\zeta_s/\zeta_c \gtrsim 4.5$) the elastic stress exerted on the thinning fluid filament becomes sufficient to influence the thinning dynamics during the early stages of polymer deformation, and thus the time scale at which elastic stresses become measurable approximately matches the longest relaxation time of the polymer, i.e., $\tau_{EC}\approx \tau$. Accordingly, the EC regime is initiated as the Weissenberg number exceeds $Wi \approx 0.5$ and within the EC regime the polymers behave as elastic Hookean springs (as per the Oldroyd-B model), continuing to stretch and accumulate strain towards near full extension when the filament breaks. In contrast, stiffer polymers with screened HI interactions reach the EC regime in a significantly stretched state because they do not exert sufficient tension to modify the Newtonian-like thinning of the fluid filament until they are nearly fully extended and already close to behaving as nonlinear springs. This leads to a reduced EC regime and a time scale for measurable elastic stress $\tau_{EC} \ll \tau$. This also means that $Wi$ can reach a very high value $\gg2/3$ prior to the onset of the EC regime. Note that the dependence of $Wi_{EC}$ on $1/L$ is shown in the insert of Fig.~\ref{fgr:Fig4}(d).

In summary, capillary thinning extensional flow techniques are widely used and appealing  due to their simplicity, minimal fluid volume requirements, and their provision of data to complement characterization by standard shear flow techniques. They also offer valuable insight into how a fluid responds to a self-selected uniaxial extensional flow that is generated in the thinning filament. However, we have shown that the longest relaxation time $\tau$ of a polymer can not be universally retrieved by extracting a characteristic time $\tau_{EC}$ from the exponential filament decay in the EC regime of a capillary thinning experiment. This limitation stems from the polymer dynamics that occur during capillary thinning, which are not universal but which strongly depend on the polymer extensibility. Our estimates indicate that for highly extensible macromolecules ($L \gtrsim 30$) the dynamics that occur during capillary thinning are in reasonable agreement with the predictions of models for which polymer orientation occurs primarily within the EC regime and which equate $\tau_{EC}$ with $\tau$. However, for macromolecules of lower extensibility it appears that significant (even complete) orientation can occur prior to the onset of the EC regime. Consequently, the duration of the EC regime is reduced prior to the onset of finite extensibility effects with the result that $\tau_{EC}$ is reduced relative to $\tau$. The criterion proposed by Campo-Dea\~no and Clasen\cite{Campo2010} is predictive of a lower concentration limit ($c_{low}$) above which the onset of polymer stretching (i.e., when the polymer behaves as an Hookean spring) generates sufficient stress to induce an EC regime. Unfortunately, for most polymers $c_{low}$ is difficult to compute except for the case of certain model polymer-solvent systems.

Previous research has highlighted inconsistencies between the characteristic time obtained from capillary thinning and that determined through traditional steady or oscillatory rheometric shear flow methods.\cite{arnolds2010capillary,sachsenheimer2014experimental,Jimenez2018} This has led to the concept that capillary thinning provides the ``extensional relaxation time'' of the polymer, a different time scale from the ``shear relaxation time'' retrieved from shear flow experiments. We acknowledge that the time scale for the onset of polymer orientation in shear and extension can be different, and that this difference is yet to be fully understood.\cite{smith1999single, Hur2002,calabrese2023extensibility} However, our experiments, for the first time, compare capillary thinning measurements against another extensional flow (i.e., that in the OSCER device). Since the time scales obtained do not (in general) agree, the idea that capillary thinning yields the extensional relaxation time cannot be completely correct. In fact, based on our understanding, we would argue that $\tau_{EC}$ should not be thought of as a relaxation time, but rather as an inverse strain rate at which elastic stresses become measurable in an extensional flow. Although, in general $\tau_{EC} \neq \tau$, we believe that $\tau_{EC}$ may in fact be a more appropriate metric to describe the elastic nature of the fluid under extensional flows, which is not necessarily captured by $\tau$ (i.e., for relatively stiff and inelastic polymers with large $\tau$ but small $\tau_{EC}$). On the other hand, when the polymer extensibility is low ($L \lesssim 30$), ``relaxation time’’ measurements made by capillary thinning techniques should be interpreted carefully as they most likely do not accurately describe the orientational dynamics of the polymer. 

This work highlights the need to develop experimental systems to properly measure the orientational dynamics of macromolecules during filament thinning, which are currently either inferred from simulations or estimated by applying significant simplifying assumptions on experimental thinning data. Such dynamical measurements may aid in the development of microstructural models and analytical expressions for the extrapolation of the true material properties of diverse complex fluids.

\section*{ACKNOWLEDGMENTS}

The authors gratefully acknowledge the support of Okinawa Institute of Science and Technology Graduate University with subsidy funding from the Cabinet Office, Government of Japan. V.C., S.J.H., also acknowledge the financial support from the Japanese Society for the Promotion of Science (JSPS, grant nos. 22K14738 and 21K03884). The authors thank Fabian Hillebrand for the stimulating discussions.

%

\end{document}